\documentclass[prl,manuscript,aps]{revtex4}

\usepackage{amssymb,amsmath}
\usepackage{graphicx,float}

\begin{document}

\title{ \Large \bf 
Derivation of power-law distributions within standard statistical mechanics}
\author{\large  \flushleft 
Rudolf Hanel$^{1,2}$  and  Stefan Thurner$^{1,}$\footnote{
 {\bf Correspondence to:}\\
 Stefan Thurner  \\Complex Systems Research Group, Medical University Vienna\\
 W\"ahringer G\"urtel 18-20; A-1090 Vienna, Austria\\
 T +43 1 40400 2099 F +43 1 40400 3332; thurner@univie.ac.at
} 
}
\affiliation{ 
$^1$ Complex Systems Research Group; HNO; Medical University Vienna; 
 Austria   \\ 
$^{2}$ Z.V.O.L. Vrouwstraat 12; B-2170 Merksem; Belgium \\
 \hspace{1cm}
} 
\begin{abstract}
\noindent
We show that within classical statistical mechanics it is possible 
to naturally derive 
power law distributions which are of Tsallis type. 
The only assumption is that microcanonical 
distributions have to be separable from of the total system 
energy, which is reasonable for any sensible measurement.  
We demonstrate that all separable distributions are parametrized 
by a separation constant $Q$ which is one to one related to 
the $q$-parameter in Tsallis distributions. 
The power-laws obtained are formally equivalent to those obtained 
by maximizing Tsallis entropy under $q$ constraints. 
We further ask why nature fixes the separation constant $Q$ to 1 
in so many cases leading to standard thermodynamics. 
We answer this with an  
explicit example where it is possible to relate $Q$ to 
sytem size and interaction parameters, characterizing the physical system. 
We argue that these results might be helpful to explain the ubiquity of 
Tsallis distributions in nature.\\


\end{abstract}
\maketitle

\section{Introduction} 
There has been a tremendous interest in a 
generalized 
definition of entropy, recently introduced by Tsallis 
\cite{tsallis88}. 
\begin{equation}
S_q=\frac{1-\int d\Gamma \rho^q }{q-1}   \quad ,
\end{equation}
where $\rho$ is the the normalized energy density and $d\Gamma$ 
indicates phase space integration. 
The reason why this modification has attracted so much interest is 
partly because of the possibility to derive power-law distributions 
in the canonical ensemble  
within the maximum entropy principle. 
This modification of entropy and its resulting formalism,
which is sometimes referred to as
non-extensive thermostatistics, 
has triggered far more than a thousand 
works in the past few years \cite{tsallis_bibliogr}. 
However, despite its phantastic descriptive success of power-laws
in physical, chemical, biological, and social systems,
it has not yet been possible to derive this form of entropy from 
thermodynamic or statistical principles. 
Within the formalism  suggested by Tsallis it is necessary to 
define expectation values of quantities depending on the 
energy spectrum  of the system not in the standard way
but with so-called $q$-expectations
$\langle {\cal O}\rangle=\int d\Gamma \, \rho^q \, {\cal O}$
\cite{tsallis,tsallis98}
to recover the Legendre structure of thermodynamics  
\cite{vives02}. 
The physical interpretation of these $q$-expectations
is under heavy debate. 

There have been several papers recently with the aim  
to derive canonical power distributions from first principles, see 
e.g.  \cite{abe,mendes01,vives03}. 
In convincing
work  \cite{almeida01,almeida02} it has been 
beautifully noted that the expression 
$\frac{d}{dE} \left( \frac{1}{\beta}\right)=q-1$ gives a 
physical meaning to $q$ and that power laws 
in the canonical ensemble can be derived on a Hamiltonian basis.  

In this work we adopt a somewhat different strategy and derive 
-- using a mathematical theorem --
power-law distributions for the canonical ensemble 
directly, just by the use of the variational principle, 
and a separation Ansatz, 
without touching the standard definition of entropy. 

\section{Power-laws in the canonical ensemble} 

We begin by noting that any thermodynamic system which can be 
measured in equilibrium must be {\it separable}, i.e.,   
that the thermodynamic  quantities of the measured system should 
not explicitely depend on the energy of the total system $E$.   
In the following we consider a sample (observed system)
in contact with a reservoir. The energy of the 
sample is $E_1$, the  energy of the reservoir is 
$E_2$, such that the total (isolated) system has a 
constant total energy
$E=E_1+E_2$. The number of microstates in the sample  
is $\omega_1(E_1)$ and $\omega_2(E_2)$ in the reservoir.
The energy of the sample fluctuates around its equilibrium (extremal)
value denoted by $E_*$. The Hamiltonians describing the 
sample and the reservoir are $H_1$ and $H_2$, respectively. 
Thermal contact of the two systems means $H=H_{1}+H_{2}$ 
and the partition function $Z(E)$ is the convolution 
of the two microcanonical densities 
\begin{equation}
Z(E) =  \int\limits^{E}_{0} dE_{1} 
                  \omega_{1}(E_{1}) \omega_{2}(E-E_{1}) \quad ,
\end{equation}
with
\begin{equation}
\omega_{i}(E_{i})=\int d\Gamma_{i}\quad\delta(H_{i}-E_{i}) \quad .
\end{equation}
Following the usual line of thought to pass from the microcanonical
to the canonical description,  represented  by $\rho$ is given  
(up to a constant multiplicative factor) by
\begin{equation}
\rho(E_{1})=\omega_{1}(E_{1})\omega_{2}(E-E_{1})Z^{-1}(E) \quad .
\end{equation}
Note, that this description is dictated by the equations of motion. 
Assuming the existence of a 
unique extremal configuration  
at some $E_{1}=E_*$ defined by
$\delta \rho=0$, leads to the well known condition
\begin{equation}
        \frac{\omega_{1}'}{\omega_{1}}\big|_{E_1=E_*}  =
	\frac{\omega_{2}'}{\omega_{2}}\big|_{E_2=E-E_*}:=
        \frac{1}{kT}= \beta  \quad , 
\end{equation}
which defines the temperature $T$ of the system. The usual 
definition of entropy $S_{i}=k\ln(\omega_{i})$ implies that 
the extremal configuration is found where $S=S_{1}+S_{2}$ 
is extremal with its associated temperature as defined above.  
 
Under which circumstances can one factorize  
the dependence of $\rho$ on the total energy $E$? 
We are hence looking for classes 
of microcanonical distributions that allow for such a 
separation of $E$ into a multiplicative 
factor. A standard way to motivate the appearance of the 
Boltzmann term in the canonical ensemble can be seen 
as a consequence of this $E$-separation  
\begin{equation}
\begin{array}{lcl}
\omega_{2}(E-E_{1})&=&\exp\bigl(\ln (\omega_{2}(E-E_{1}))\bigr)\\
&\approx& \exp\bigl(\ln(\omega_{2}(E))-
	\frac{\partial}{\partial E}\ln(\omega_{2}) \, E_{1} \bigr)\\
&\approx& \omega_{2}(E)\exp(-\beta E_{1}) \quad .
\end{array}
\label{boltzfact}
\end{equation}
It is worthwhile to note that the approximation in  Eq. 
(\ref{boltzfact}) is exact for $\omega_{2}(E-E_{1})$ being 
an exponential in $E$.  Up to this point we have summarized 
textbook knowledge.  
It is one purpose of this work to emphasize  
that (\ref{boltzfact}) is not the most general 
way of separation.  

To find the most general separation, we generalize the log function in Eq.
(\ref{boltzfact}) to some real function $f$, being strictly 
monotonous and twice differentiable. 
Monotonicity is needed for a well defined inverse $f^{-1}$.
The idea is to write  
$\omega_{}(E-E_{1})=f^{-1}\circ f \circ \omega_{}
  \bigl((E-E_*)-(E_{1}-E_*)\bigr)$
and to expand $f\circ\omega_{}$ around $E-E_*$.
Suppose energy $E$ is separable from the system, then  
there exist two functions $g$ and $h$ such that 
\begin{equation}
   \omega_{}(E-E_{1})=g\bigl(\omega_{}(E-E_{*})\bigr) \,\, h(x)  \quad , 
\label{seperab}
\end{equation}
with $x:=\beta (E_{1}-E_{*})$; to simplify notation   
we write $\bar \omega:= \omega(E-E_{*})$ in the following.
We now use $f$ to find the unknown functions $g$ and $h$ by  
expanding $f\circ\omega$ to first order 
\begin{equation}
 f\bigl(\omega(E-E_1)\bigr)  
 = f\bigl(g(\bar \omega) h(x)\bigr)
 \sim   f(\bar \omega)-\bar \omega \, x \, f'(\bar \omega) \quad , 
\label{sep}
\end{equation}
which is justified for small $x$, i.e., the system being 
near equilibrium. 
The most general solution to this separation Ansatz is 
given by the family of equations $(f,g,h)_Q$, 
parametrized by a separation constant $Q$, 
and $C$ and $C_2$ being real constants  
\begin{equation}
 \begin{array}{lcl}
 f_{}(\omega)&=&C\, \omega^{1-Q} +C_2   \\
 g_{}(\omega)&=&\omega    \\
 h_{}(x)     &=&\bigl[1-(1-Q)x \bigr]^{\frac{1}{1-Q}}   
\end{array}
\label{result1}
\end{equation}
To see this, first $g$ is found by setting $x=0$ and $h_0=h(0)$, 
so that Eq. (\ref{sep}) yields $f(g(\bar \omega) h_0)= f(\bar \omega)$, 
which means $g(\bar \omega) =\frac{\bar \omega}{h_0}$. 
Without loss of generality set $h_0=1$ and arrive at 
$f(\bar \omega h(x)) = f(\bar \omega) - \bar \omega x f'(\bar \omega)$. 
Form partial derivatives of this expression with respect 
to $x$ and $\bar \omega$, and eliminate the $f'(\bar \omega h)$ term  
from the two resulting equations 
\begin{equation}
\begin{array}{lcl}
f'(\bar \omega h) h'   &=& - f'(\bar \omega)  \\
f'(\bar \omega h) h    &=&  (1-x)f' -\bar \omega x f'' 
\end{array} 
\end{equation}
to  arrive at the separation equation 
\begin{equation}
1-\frac{1}{x}\left( \frac{h}{h'} +1\right)  = 
- \bar \omega \frac{f''(\bar \omega)}{f'(\bar \omega)} =Q 
\end{equation}
where $Q$ is the separation constant.  The differential 
equation $1-\frac{1}{x}\bigl( \frac{h}{h'} +1\bigr)  =Q $
is straight forwardly solved to give 
$h(x)=\bigl[1-(1-Q)x\bigr] ^{\frac{1}{1-Q}}$, using $h(0)=1$ 
to fix the integration constant. 
The equation $- \bar \omega \frac{f''(\bar \omega)}{f'(\bar \omega)} =Q$
means, 
$f(\bar \omega) =C_1 \frac{1}{1-Q} \bar \omega^{1-Q} +C_2$, with 
$C_1$ and $C_2$ integration constants. $f$ is 
strictly monotonous except for $Q=1$, where it is constant.  
It is straight forward to test that Eqs. 
(\ref{result1}) solve  Eq. (\ref{sep}). 
The term of interest in the canonical distribution can now be written  
as a generalized Boltzmann factor 
\begin{equation}
\omega_{2}(E-E_{1})= 
 \omega_{2}(E-E_*) \bigl[ 1-(1-Q) \beta(E_1-E_*) \bigr]^{\frac{1}{1-Q}}
\quad .
\label{qdistrib}
\end{equation}
The separation constant is not specified at this level. 
As we will see below the choice of a particular physical 
system will determine $Q$.   
The usual Boltzmann factor Eq. (\ref{boltzfact}) is recovered 
as a special case in the limit $Q\to 1$.
Note, that if $\omega_{2}$ is of the form $\omega_{2} \propto E^{1/1-Q}$, 
then Eq. (\ref{qdistrib}) holds exactly and not only to the 
first order approximation in Eq. (\ref{sep}). The best way to 
prove this is to write 
 $\beta= \frac{\omega_{2}'}{\omega_{2}}|_{E-E_{*}}= \frac{1}{(1-Q)(E-E*)}$
and to compute straight forwardly
\begin{equation}
\begin{array}{lcl}
\omega_{2}(E-E_{1}) =
 (E-E_{*})^{\frac{1}{1-Q}}
 \bigl(1-\frac{E_{1}-E_{*}}{E-E_{*}}\bigr)^{\frac{1}{1-Q}} \\
 =\omega_{2}(E-E_{*})\bigl[1-
 (1-Q)\beta(E_{1}-E_{*})\bigr]^{\frac{1}{1-Q}}  \quad , 
\end{array}
\end{equation}
see also \cite{plastino}. 
As we will see below, $\omega_{2} \propto E^{1/1-Q}$ 
covers classical (homogenous) Hamiltonians with pair-potentials. 

Having derived the principal form  $(f,g,h)_{Q}$ to first order 
it is easy to see that with the same family of functions
we can expand $\omega_{2}$ to all orders. 
Repeated differentiation of $\rho$ at its extremum 
leads to a hierarchy of equations relating properties 
of the $\omega$ densities to properties of $\rho$ at 
its extremum. With the definitions
\begin{equation}
  r_{n}:=\beta^{-n} \, \frac{\rho^{[n]}}{\rho} \, \big|_{E=E_{*} }
  \quad {\rm and} \quad 
  \phi_{n}^i:=\beta^{1-n}  \, \frac{\omega_i^{[n]}}{\omega_i^{[1]}}
  \big|_{E=E_{i*} }
\end{equation}
$[n]$ being the $n$ th derivative, and $i=1,2$ indicating 
system 1 and 2 ($E_{1*}=E_*$, $E_{2*}=E-E_*$). 
We successively construct the whole hierarchy to find
\begin{equation}
\begin{array}{lcl}
r_2 &=&  \phi_2^1 +\phi_2^2 -2\\
r_3 &=& (\phi_3^1-3\phi_2^1) - (\phi_3^2-3\phi_2^2)\\
r_4 &=& \bigl(\phi_4^1-4\phi_3^1 +3\phi_2^1(r_2 +2 -\phi_2^1)\bigr) \\
    &+&         \bigl(\phi_4^2-4\phi_3^2 +3\phi_2^2(r_2 +2 -\phi_2^2)\bigr) \\
 &\vdots&
\end{array}
\end{equation}
Using these equations we can re-express the $\phi_n^2$ terms in terms 
of $\phi_n^1$ and $r_n$, so that all we know about system 2 is 
encoded in local properties of $\rho$ at the equilibrium.
To fourth order the general expansion reads  
\begin{equation}
\begin{array}{lcl}
\omega(E&-&E_{1})= \omega(E-E_{*})\bigl\{ 1+(1-Q) \bigl[-x \\
  &+& \frac{1}{2!}(\phi_2^2-Q)\, x^{2} 
  -\frac{1}{3!}\bigl(-3Q\phi_2^2 + \phi_3^2 +(Q+1)Q\bigr)\, x^{3} \\ 
  &+&\frac{1}{4!}
    \bigl(\phi_4^2 - 4Q\phi_3^2-3Q(\phi_2^2)^2
    + 6Q(Q+1)\phi_2^2 \\
   &-&(Q+2)(Q+1)Q\bigr)\, x^{4} 
  + \dots \bigr] \bigr\}^{\frac{1}{1-Q}} \quad ,
\end{array}
\end{equation}
which in the limit $Q \rightarrow 1$ is 
\begin{equation}
\begin{array}{lcl}
\omega(E&-&E_{1})=\omega(E-E_{*}) \exp \bigl\lbrace - x + 
  \frac{1}{2!}\bigl( \phi_2^2 -1\bigr)x^{2} \\ 
  &-& \frac{1}{3!}\bigl(-3\phi_2^2 + \phi_3^2 + 2\bigr)x^{3}  
  +\frac{1}{4!}
  \bigl(\phi_4^2 - 4\phi_3^2-3(\phi_2^2)^2 \\
  &+& 12\phi_2^2 -6\bigr)x^{4}+ 
  \dots \bigr\rbrace
\end{array}
\label{order4}
\end{equation}
Note, that for $\omega_2\propto\exp(\beta E)$ all 
$\phi_n^2=1$, and all higher order terms vanish and the 
standard Boltzmann result becomes exact.  
This concludes the main finding of the paper. 

\section{The physical meaning of the separation constant Q}
Why does nature fix $Q\longrightarrow 1$  in so many cases, 
i.e. why is standard thermodynamics the most predominantly realized 
situation?  
In the following 
we will demonstrate with the help of an example how the 
separation constant $Q$ appearing in  Eq. (\ref{qdistrib}) can be 
related to system size and interaction parameters of a real 
physical system. This explains the ubiquity of $Q=1$. Examples 
of this kind have been given in a somewhat different context 
before \cite{tsallis,almeida02}.  

\subsection{An example}
Let us specify the following $N$ particle 
Hamiltonian for pair-potentials, governing the  sample
in $D$ space dimensions. We use $n=DN$. 
\begin{equation}
H(x,p)= \sum_{i}^{N} \frac{p_i^2}{2m} + \sum_{i<j}^{N} |x_i-x_j|^{\alpha} 
\quad . 
\end{equation}
The energy density is given by the phase space integral 
\begin{equation}
\begin{array}{lcl}
\omega (E) & =&  \int d^np \, d^nx \,\, 
        \delta\left(\sum_{i}  \frac{p_i^2}{2m} 
                   +\sum_{i<j} |x_i-x_j|^{\alpha} -E \right)  \\
 & =& \int_{0}^{E} dE_1  \int d^np \, d^nx \,\,
     \delta\left(\sum_{i} \frac{p_i^2}{2m} -E_1 \right) 
     \\& \times &
     \delta \bigl(\sum_{i<j} |x_i-x_j|^{\alpha}-(E-E_1)\bigr) \quad . 
\end{array}
\label{zwi}
\end{equation}
We compute the kinetic term  
\begin{equation}
\begin{array}{lcl}
 & \int& d^np \,\, \delta\left( 
    \sum_{i=1}^N \frac{|\vec p|^2}{2m} -E \right) 
  = \int_{\frac{\vec p^2}{2m} = E} d {\cal O}_n 
           \left|\vec \bigtriangledown \frac{\vec p^2}{2m}\right|^{-1} \\
  & =&  \int_{|\vec p| = \sqrt{2mE}} d {\cal O}_n \frac{m}{|\vec p|} 
  = \sqrt{\frac{m}{2E}}  \int_{|\vec p| = \sqrt{2mE}} d {\cal O}_n \\
  & =& {\cal O}_n  \sqrt{\frac{m}{2E}}  \bigl(2mE\bigr)^{\frac{n-1}{2}} 
  \propto E^{\frac{n}{2}-1}
\end{array}
\end{equation}
and the potential contribution
\begin{equation}
\begin{array}{lcl}
 &\int& d^n x \,\, \delta \bigl( 
     \sum_{i<j} |x_i-x_j|^{\alpha}-E \bigr) \\
 &=& \int d^n x \,\, \delta \bigl( 
     \sum_{j=2}^N \sum_{i=1}^{j-1} 
     \left| \sum_{k=1}^{D}(x_i^k-x_j^k)^2 \right| ^{\frac{a}{2}}
     -E \bigr) \\
 &=& \int d^n x \,\, \delta \bigl( E \left[ 
     \sum_{j=2}^N \sum_{i=1}^{j-1} 
     \left| \sum_{k=1}^{D} \left( \frac{x_i^k-x_j^k}
     { E^{\frac{1}{a}}} \right) ^2 \right| ^{\frac{a}{2}} -1 
      \right] \bigr) \\
 &=&  E^{\frac{n}{a}} \int d^n y \,\, \delta \bigl( E \left[ 
      \sum_{j=2}^N \sum_{i=1}^{j-1} 
      \left| \sum_{k=1}^{D} \left(y_i^k-y_j^k\right) 
      ^2 \right| ^{\frac{a}{2}} -1 
      \right] \bigr) \\
&=&  E^{\frac{n}{a}-1} \cdot  const.
\end{array}
\end{equation}
where we used the substitution $y_i=x_i/E^{\frac{1}{a}}$  and the 
fact $\int dx \, \delta(\lambda x)= \int dx \, \lambda^{-1}\delta(x)$.  
We finally get for Eq. (\ref{zwi}) 
\begin{equation}
 \omega(E) \propto   \int_0^E dE_1 \,\,  E_1^{\frac{n}{2}-1} \,\,
            \bigl(E-E_1\bigr)^{\frac{n}{\alpha}-1}
           \propto  E^{\frac{(\alpha +2)n}{2\alpha}-1} \quad .
\label{hamilton}
\end{equation}
This allows us now to compare exponents (and coefficients) in 
Eq. (\ref{hamilton}) and Eq. (\ref{qdistrib}) to arrive at the 
relation 
\begin{equation}
\frac{1}{1-Q}=\frac{(\alpha+2)n}{2 \alpha}-1 \quad  , 
\label{relation}
\end{equation}
which fixes the separation constant. 
This equation establishes the connection between the interaction  
term in the Hamiltonian $\alpha$, the dimensionality of the phase 
space $n=DN$, and the separation constant $Q$. 
From Eq. (\ref{relation}) it is immediately clear that 
for large systems the separation constant is always 
$Q\to 1$, i.e. the classical Boltzmann term (\ref{boltzfact}) 
is recovered. For small systems, with a fixed number of particles
$Q$ depends on the interaction between the particles. 
For an ideal gas $\alpha \to -\infty$ the separation constant is 
$Q=\frac{4-n}{2-n}$. 
Nontrivial $Q\neq1$  should be expected 
for strongly interacting and/or small systems, i.e., 
$|\alpha|/|\alpha+2| \sim  n$
and systems with $-2<\alpha<0$, where the limit $n\rightarrow\infty$ implies
$n(\alpha+2)/(2\alpha)-1\rightarrow-\infty$ so that BG is not obtained.
Let us assume that for a system $Q\neq1$ is due to its 
finite size. Here the standard argument in BG thermodynamics 
of  extremely sharp peaks of the distribution at equilibrium 
is not necessarily valid, and $E_*$ and the expected energy 
${\cal U}=\langle H \rangle$ may significantly differ. Assuming 
$\omega_i(E_i)\propto E_i^{n_i}$ and the validity of the 
variational principle  
$\delta \rho =(\frac{n_1}{E_1}-\frac{n_2}{E-E_1}) \rho \delta E_1=0$
yields
$E_1=E_*=\frac{E}{1+\frac{n_2}{n_1}}$. 
With this we can write 
$\int_0^EdE_1 \, E_1^A(E-E_1)^B = 
\left( \frac{E_*}{n_1}\right)^{A+B+1} 
\int_{-n_1}^{n_2} dx (n_1+x)^A(n_2-x)^B$
and compute 
\begin{equation}
{\cal U}= \langle H_1\rangle_E= \frac{\int_0^EdE_1\, E_1 \rho(E_1)}
                                     {\int_0^EdE_1\,  \rho(E_1)}=
E_*\left( 1+ \frac{\langle x \rangle_{12}}{n_1}\right)
\end{equation}
with 
$\langle x \rangle_{12}=\frac{\int_{-n_1}^{n_2} dx x (n_1+x)^{n_1}(n_2-x)^{n_2}}
               {\int_{-n_1}^{n_2} dx   (n_1+x)^{n_1}(n_2-x)^{n_2}}$.
Rewriting leads to  
\begin{equation}
{\cal U}= \frac{E}{1+\frac{n_2-\langle x \rangle_{12}}
                          {n_1+\langle x \rangle_{12}} } \quad . 
\end{equation}
One can therefore apply the same equilibrium argumentation 
also to the (measurable) expectation values by substituting 
$\tilde n_1= n_1+\langle x \rangle_{12}$, 
$\tilde n_2= n_2-\langle x \rangle_{12}$ and 
$\tilde \rho \propto E_1^{\tilde n_1} (E-E_1)^{\tilde n_2}$ 
so that $\delta \tilde \rho=0$ with $\delta E=0$. 
The $Q$ of the observable canonical ensemble is then 
\begin{equation}
\frac{1}{1-\tilde Q}=\tilde n_2 \quad  , 
\label{relation2}
\end{equation}
for the example of a pair potential Hamiltonian 
for the reservoir.

\section{Conclusion}

Equation (\ref{qdistrib}) is without doubt formally 
equivalent to the Tsallis $q$ form of the Boltzmann term. 
Our result (\ref{qdistrib}) appears as a consequence of 
the separability of the total system
energy $E$ from the microcanonical density of the sample  
system with the parameter $Q$ being nothing but a separation 
constant not further specified at this stage. 
$Q$ is fixed and a physical meaning is 
obtained as soon as a particular Hamiltonian is specified. 
Equation (\ref{qdistrib}) is not exactly what 
Tsallis gets by extremization of $S_q$, with 
his constraints for the canonical ensemble: $\sum_i p_i=1$, and 
$U_q=\frac{\sum_i p_i^q \epsilon_i}{\sum_i p_i^q} = const.$
In our derivation we get $E_*$ and the classical 
$\beta$ (real temperatures),  where Tsallis gets the terms $U_q$ and 
$\beta/\sum_i p_i^q$, respectively. 
Our general Boltzmann factor is obviously not obtained as a 
result of a maximization of $k\sum_i p_i\ln p_i$. 

We have computed higher order terms and found that 
if the canonical distribution is a power, all higher 
terms in the expansion vanish and our result (\ref{qdistrib})
holds exactly and not only as a first order approximation
as we started out with in (\ref{sep}). 
This might be an interesting finding since many physically 
interesting microcanonic densities behave like powers 
in $E_1$, at least for finite size systems. 

An important aspect in our work is that all our 
arguments are strictly based on Hamiltonians and on 
the variational  principle. At no point we are 
forced to take the thermodynamic limit. 
For that case we had some discussion if the equilibrium 
point is different from the expectation value. 
In case that mean and equilibrium do not coincide, our result would 
of course not be observable, since measurements take place 
at the expectations whereas we have developed our results 
around the extremal configuration.  We can however show that for 
microcanonical distributions characterized by 
$\omega_1(E_1) \propto  E^{\kappa_1}$ and 
$\omega_2(E_2) \propto  E^{\kappa_2}$, mean and extremum coincide 
for $\kappa_1=\kappa_2$. 
In the thermodynamic limit the difference is of course completely 
irrelevant. 

The purpose of the given example is not to relate 
the first part of the paper to any non-extensive system,
but only to demonstrate the possibility to 
relate the separation constant to physical parameters. 
However, we think that it is in principle possible to use the 
central part of the paper to think about the occurrence 
of phenomena in situations described in e.g. \cite{rapisarda}.
Here within the long meta-stable regions we have the situation of 
having an 'almost-equilibrium' which  might be characterized by 
a non-trivial $Q$  before the $t\to \infty$ limit is taken.

We thank H. Grosse for his interest  
and C. Tsallis for useful comments.

\newpage 
 
\bibliographystyle{unsrt}

\end{document}